# THE MECHANICAL AND ELECTRICAL PROPERTIES OF DIRECT-SPUN CARBON NANOTUBE MAT-EPOXY COMPOSITES

Wei Tan, Joe C. Stallard, Fiona R. Smail, Adam M. Boies, Norman A. Fleck*

Engineering Department, University of Cambridge, Trumpington Street, Cambridge, CB2 1PZ, UK.

*Corresponding author. Tel: +44 (0)1223 748240. Email: naf1@eng.cam.ac.uk (N.A. Fleck)

## Abstract:

Composites of direct-spun carbon nanotube (CNT) mats and epoxy are manufactured and tested in order to determine their mechanical and electrical properties. The mats are spun directly from a floating catalyst, chemical vapour deposition reactor. The volume fraction of epoxy is varied widely by suitable dilution of the epoxy resin with acetone. Subsequent evaporation of the acetone, followed by a cure cycle, leads to composites of varying volume fraction of CNT, epoxy and air. The modulus, strength, electrical conductivity and piezoresistivity of the composites are measured. The CNT mats and their composites exhibit an elastic-plastic stress-strain response under uniaxial tensile loading, and the degree of anisotropy is assessed by testing specimens in 0°, 45° and 90° directions with respect to the draw direction of mat manufacture. The electrical conductivity scales linearly with CNT volume fraction, irrespective of epoxy volume fraction. In contrast, the modulus and strength depend upon both CNT and epoxy volume fractions in a non-linear manner. The macroscopic moduli of the CNT mat-epoxy composites are far below the Voigt bound based on the modulus of CNT walls and epoxy. A micromechanical model is proposed to relate the macroscopic modulus and yield strength of a CNT mat-epoxy composite to the microstructure.

# Graphical abstract

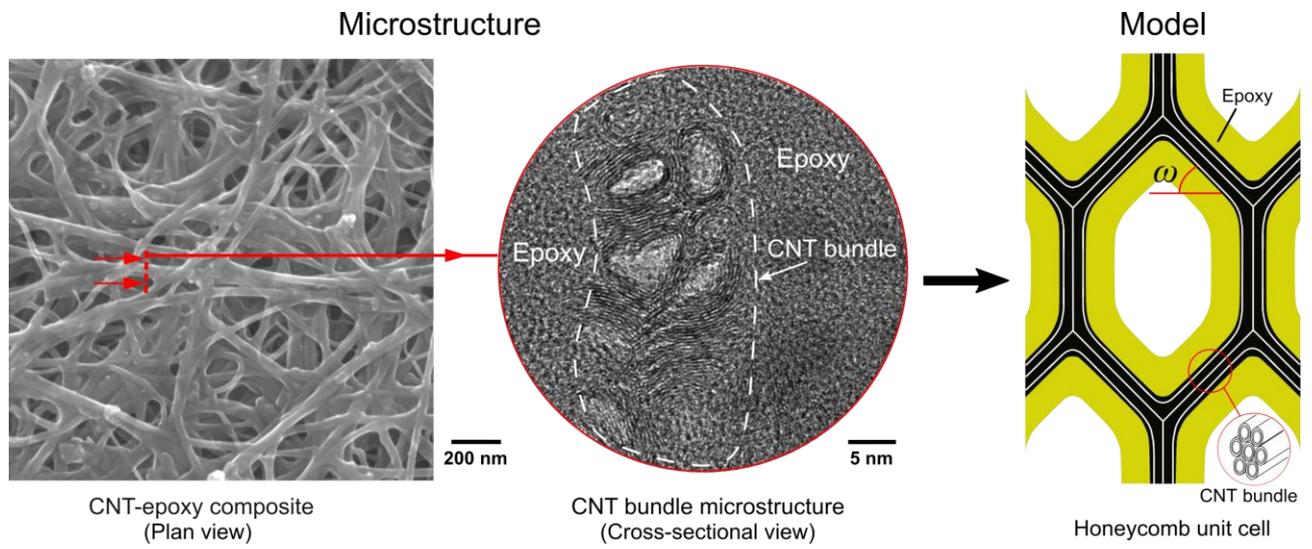

# 1. Introduction

## 1.1 Properties of CNT-polymer composites

The impressive mechanical, electrical and thermal properties of individual carbon nanotubes (CNTs) have attracted considerable interest from the research community [1,2]. The wall of a CNT has an elastic modulus of about 1 TPa, a tensile strength of 100 GPa [3], a thermal conductivity of 3500 W/mK [4], and an electrical conductivity of $2 \times 10^7$ S/m [5]. Several thousand tonnes of CNTs are now produced each year in powder form [1], yet the annual production of CNT mats and yarns is only a small proportion of global output. Although the high production cost precludes their use in many engineering applications, it is anticipated that CNT mats will be produced in industrial volumes, suitable for processing into high-performance CNT-based composites [6].

CNT-polymer composites exist in five categories, as labelled (a) to (e) in Figure 1, and are summarised as follows:

*Class (a)*: Short CNTs in powder form can be dispersed in molten polymers or polymer-solvent solutions, often with the aid of surfactants [7]. These CNT-polymer solutions can then be cast into a mould to form CNT-polymer composites [8–10]; the content of CNTs in such composites is typically less than 4% by weight.

*Classes (b) and (c)*: Alternatively, CNT-solvent solutions can be injected into a bath of coagulating polymer solution to form unidirectional CNT-polymer fibres [11–13], termed *class (b),* or filtered to create planar isotropic 'buckypaper' mats, which are subsequently processed into planar composites via the infiltration of polymer solutions [14–17], *class (c).*

*Class (d)*: Vertically aligned CNT forests are grown from substrates. They can then be infiltrated with polymers and cured in the vertically aligned form [18], or processed into a mat in the dry state by drawing or by flattening via the so-called 'domino pushing' process [19–21]. In a variant of this class, twisting is imposed during drawing or after flattening to produce a unidirectional fibre. Mats or fibres manufactured from CNT forests are subsequently processed into CNT-polymer composites via the infiltration and cure of polymer solutions [20–29].

*Class (e)*: CNT mats and fibres are collected continuously by drawing CNT aerogels from a floating catalyst chemical vapour deposition reactor. These so-called "direct-spun" materials can be infiltrated with polymer solutions to create CNT-polymer composites [30–35]. It is this last set of materials which form the focus of the current study. The performance of such composites in the drawn direction may be enhanced by stretching [30,31] or by the application of pressure during cure [36]: these processes can raise the CNT volume fraction and align the CNTs with the loading direction.

The reported moduli and strength of these classes of CNT-based composites [8-17,20-35] are assembled in Figure 2(a). For comparison, the strength and modulus of individual CNT walls [3,37] are included. Note that the moduli and strength of CNT-based composites span several orders of magnitude for composite densities in the narrow range of 800 kg/m$^3$ to 1200 kg/m$^3$. Also, the moduli and strength of CNT composites are much below those of CNT walls [38–43], and are dependent on the orientation and waviness of the CNT reinforcement [44], and on the CNT volume fraction [18]. The chart of electrical and thermal conductivity Figure 2(b), assembled from the data of [4,17,27,45–54], reveals that the electrical and thermal conductivity of CNT composites are also orders of magnitude below those of individual CNT walls, and these properties vary not only within a class of material but also between composite classes.

In broad terms, CNT-epoxy composites consist of a combination of three phases: CNTs, epoxy and air. A composition diagram is given in Figure 2(c) to indicate the relative volume fraction of CNT bundles $f_B$, epoxy $f_e$ and of air $f_a$. The figure includes CNT foams [55], CNT fibres [56,57], and CNT mats [57–60] for which $f_e = 0$. CNT foams possess a CNT volume fraction $f_B < 0.10$, whilst dry CNT fibres and mats possess $0.10 < f_B < 0.82$. Most quasi-isotropic CNT-epoxy composites [33,35,53,61] have a volume fraction $f_B < 0.40$, whereas, for aligned CNT-epoxy composites, $f_B$ is typically in the range $0.60 < f_B < 0.70$ [30,31,36].

The in-plane modulus of carbon nanotube composites $E$, as reported in [8-17,20-35] and as measured in this study, are compared with those of long fibre carbon fibre composites in Figure 2(d) by plotting $E$ versus reinforcement volume fraction $f_R$. For CNT bundles, we take $f_R = f_B$. The normalisation of $E$ for CNT-based composites and carbon fibre composites is chosen as follows. For the CNT-based composites, the modulus $E$ is normalised by that of an individual CNT bundle, $E_R$ = 680 GPa [57]. Recall that the axial modulus of a unidirectional (UD) carbon fibre-epoxy lamina can be two orders of magnitude greater than the transverse modulus due to the extreme anisotropy of the fibres. Here, we shall define the term 'composite modulus', $E$, as the largest principal value of the modulus tensor. It is this modulus $E$, normalised by the modulus of the carbon fibres as used in each separate study [62–67], that is plotted Figure 2(d) as a function of reinforcement volume fraction $f_R$. The data used for these composites are given in full in Table S1 of the supplementary information.

Voigt and Reuss bounds are included in Figure 2(d). The knockdown in macroscopic modulus from the Voigt bound is sensitive to composite microstructure. The modulus of unidirectional (UD) CFRP composites is close to the Voigt bound, whilst the modulus of a quasi-isotropic planar CFRP laminate lies below it by a factor of about ×3 due to the presence of reinforcement fibres orientated away from the principal direction. Note that the modulus of the unidirectional forest-drawn CNT-polymer composites are below the Voigt bound by more than a factor of 2, whilst those measured from composites of direct-spun CNT mats in this study lie on average more than a factor of 3 below those drawn from forests. The effect of CNT waviness, aspect ratio, and the uniformity of reinforcement volume fraction have been explored theoretically for CNT composites using classical bounds [68] and micromechanical models [44]: the presence of waviness, in particular, knocks down the macroscopic modulus by more than an order of magnitude [44].

## 1.2 Scope of Study

The purpose of the present study is to determine experimentally the effect of epoxy infiltration upon the tensile stress-strain response and electrical conductivity of direct-spun CNT mat-epoxy composites over a broad compositional range, and to understand the relationship between microstructure and properties through experiment and micromechanical modelling. Composites of CNT volume fraction between 0.11 and 0.35, and epoxy volume fraction between 0.01 and 0.79, were produced by the solvent-assisted infiltration of a direct-spun CNT mat with epoxy. The composition was controlled via the initial concentration of epoxy in acetone, and by applying varying consolidation pressures. The in-plane uniaxial tensile response and electrical properties were measured by tensile tests and a four-point electrical probe method, respectively. Scanning and transmission electron microscopy were used to characterise the epoxy distribution at the microstructural level. A micromechanical model was then developed to relate the macroscopic modulus and strength to the microstructure of the direct-spun CNT mat and CNT-epoxy composites.

## 2. Manufacture and composition of materials

### 2.1 Manufacture of CNT mat

Direct-spun CNT mat was produced by the floating-catalyst chemical vapour deposition (FCCVD) process[1], as initially described by Li et al [69]; the process is discussed in more detail elsewhere [70]. In brief, a CNT aerogel sock is drawn continuously from a FCCVD reactor and wound onto a mandrel, thereby forming a multi-layer carbon nanotube mat [57]. The hierarchical microstructure of the direct-spun CNT mat is sketched in Figure 3. The mat is of width 0.9 m and thickness 70 µm, and is comprised of flattened CNT aerogel socks. Each flattened sock is of width 80 mm and thickness about 170 nm, and is comprised of a network of branched CNT bundles, forming an interconnected network. Each bundle consists of between 10 and 40 closely packed CNTs. Macroscopically, direct-spun mats possess in-plane anisotropy in their mechanical and electrical properties, and the anisotropy is sensitive to the ratio of the velocity of the drawn aerogel to that of the gas flow within the reactor [71]. The principal material orientation is along the draw direction of the CNT sock in the reactor.

### 2.2 Manufacture of CNT-epoxy composites

The production method for fabricating the CNT-epoxy composites is sketched in Figure 4. Samples were cut from a single ply of direct-spun CNT mat of width 10 mm, length 80 mm, and nominal thickness 70 µm. They were then infiltrated by a solution of epoxy in acetone, followed by evaporation of the acetone under vacuum, and by curing in a hot press. The epoxy resin was bisphenol-A IN2 infusion resin and a hexane hardener[2]. The epoxy resin

---

[1] Provided by Tortech Nano Fibers Ltd, Hanassi Herzog St., Koren Industrial park, Ma'alot Tarshiha, 24952 Israel.
[2] IN2 epoxy infusion resin and a vacuum bagging system were obtained from Easy Composites Ltd, Park Hall Business Village, Longton, Stoke on Trent, Staffordshire, ST3 5XA, UK

and hardener were mixed in the proportion 10:3 by mass, and were then diluted with acetone. The mass concentration of resin and hardener in this solution, as denoted by $\phi$, varied from 0.001 to 1. By infiltrating the CNT mat with epoxy-acetone solutions of selected values of $\phi$, the volume fraction of epoxy in the cured composite could be suitably controlled. After immersion of the CNT mat samples in the epoxy-acetone solution for 60 s, manufacture proceeded along one of two routes, see Figure 4. In the first route, the infiltrated CNT mat samples were placed between an inner layer of PTFE release film and an outer breather layer, and a vacuum bagging system[2] was used to apply a consolidation pressure of $P = 0.1$ MPa (one atmosphere), at a temperature of 40 °C for one hour. Excess resin was absorbed by the breather layer material on each side of the CNT-epoxy composites during consolidation. Samples were then cured in a hot-press at 120 °C for 3 hours, with a constant pressure of 0.6 MPa in the through-thickness direction. An alternative production route was used to manufacture CNT composites of higher CNT volume fraction: infiltrated CNT mat samples were placed between PTFE films, and were subjected to a through-thickness pressure of $P = 10$ MPa between the loading platens of a screw-driven test machine. Since the samples were compressed between the PTFE films, squeeze-out of excess epoxy was prevented. After the consolidation pressure had been applied for three hours, allowing the epoxy to cure partially within the densified CNT mat, the samples were placed within the same vacuum bagging system as that described above (pressure of 0.1 MPa and a temperature of 40 °C for one hour) to remove residual acetone or volatile compounds. The samples were then cured in a hot press at 120 °C for three hours, under the same conditions as for the first route.

Dogbone specimens of pure, void free epoxy were also prepared for uniaxial testing. Epoxy solutions were cast into a mould, and degassed in a vacuum chamber for 1 hour, before curing at 120 °C for 3 hours. To determine the effect of acetone dilution upon the mechanical properties of the epoxy, additional dogbone samples were cast from a solution of acetone and epoxy, with $\phi$ = 0.5. The acetone was then removed by degassing under vacuum in the same manner, and the sample was cured as above.

## 2.3 Composition and physical properties of CNT mat and composites

The volume fraction of CNT bundles $f_B$ and volume fraction of epoxy $f_e$ are related to the corresponding weight fractions of CNT bundles and epoxy $w_B$ and $w_e$ respectively, as follows. Write $\rho_B$ as the CNT bundle density and $\rho_e$ as the measured epoxy density. Then, upon writing $\rho$ as the measured composite density, we have

$$f_B = w_B \frac{\rho}{\rho_B}, \qquad f_e = w_e \frac{\rho}{\rho_e}. \qquad (1)$$

The weight fraction of CNT bundles in the composite $w_B$ is the ratio of the mass of CNT bundles to the mass of the composite, and the weight fraction of epoxy follows directly as $w_e = 1 - w_B$. Both the mass of the CNT mat before infiltration and the mass of the cured CNT-epoxy composite were measured with a mass balance. The width and length of samples was measured with a Vernier scale, and the thickness was measured by micrometer, and averaged over ten readings along the sample length. The density of the composite $\rho$ was determined via its mass and volume. The dependence of thickness $t$ upon composite density $\rho$ is recorded in Figure S1 of the supplementary information. For samples of direct-spun CNT mat, the thickness was confirmed by X-ray tomography, as this method avoids compression of the sample during measurement.

The CNT bundle density $\rho_B$ was determined by helium pycnometry[3] [57], as follows. A CNT mat sample was placed in a vacuum chamber of known volume, and the chamber was then filled with helium gas. After measuring the pressure within the chamber, a valve was opened to link it to a second vacuum chamber of known volume. The final pressure was recorded after equilibrium had been attained. The ideal gas law was used to calculate the sample volume from the measured gas pressures and the known chamber volumes. The bundle density follows immediately from the sample mass divided by the sample volume, giving $\rho_B$ = 1560 kg/m³. The density of the epoxy, as calculated from the measured epoxy dogbone sample dimensions and mass, was $\rho_e$ = 1120 kg/m³. Finally, the porosity of the composites, referred to here as the volume fraction of air $f_a$, was determined by subtracting the calculated volume fraction of CNT bundles and of epoxy from unity.

---

[3] Quantachrome UK Ltd, Units 6 7 Pale Lane Farm, Pale Lane, Hook, RG27 8DH, UK

The densities of the mat and of the CNT-epoxy composites, and the volume fractions of CNT bundles, epoxy and air are recorded in Table 1 as functions of the weight fraction of epoxy in the infiltration solution $\phi$ and of the through-thickness pressure $P$ applied after infiltration. The as-received CNT mat specimens (absent infiltration) are labelled (1). CNT mat-epoxy composites manufactured at $P$ = 0.1 MPa are labelled (2) to (9) in order of increasing $\phi$, such that (9) corresponds to a composite made from undiluted epoxy ($\phi$ = 1) and CNT mat. Cured epoxy dogbone samples with initial concentration of epoxy in acetone $\phi$ = 0.5 and $\phi$ = 1 are labelled (10) and (11), respectively. Two composites (5h) and (6h) were manufactured with the high pressure $P$ = 10 MPa, from an epoxy-acetone mix equal to that of (5) and (6), respectively. Additionally, a sample of direct-spun mat (1h) was infiltrated with acetone and subjected to the same through-thickness pressure of $P$ = 10 MPa.

The dependence of as-cured density $\rho$ upon the initial epoxy concentration $\phi$ of the epoxy-acetone solution is plotted in Figure 5(a) for $P = 0.1$ MPa and $P = 10$ MPa. The density $\rho$ relates to the densities and volume fractions of CNT bundles and epoxy according to $\rho = f_B \rho_B + f_e \rho_e$. We note that $\rho$ increases monotonically with increasing $\phi$, for each selected value of $P$. Thus, it is convenient to report the composition and properties of each composite against $\rho$ rather than the process variable $\phi$. We also note from Figure 5(a) that $\rho$ increases with increasing $P$ at any fixed value of $\phi$. The volume fractions of CNT bundles, epoxy and air are plotted against the macroscopic composite density in Figure 5(b) for composites fabricated at $P = 0.1$ MPa, that is by consolidation under vacuum before curing, and in Figure 5(c) for samples that were subjected to $P = 10$ MPa pressure after infiltration. Samples fabricated at $P = 0.1$ MPa exhibit a maximum CNT volume fraction of $f_B = 0.25$ at $\rho = 771$ kg/m³, see sample (6) in Figure 5(b). In contrast, $f_B$ rises monotonically with increasing $\rho$ for the choice $P = 10$ MPa. The non-linear dependence of $f_B$ upon $(\rho, P)$ arises from the transient consolidation response of the bundles in the presence of an evaporating solvent, acetone, and a curing matrix, epoxy.

## 3. Measured properties of CNT mat and CNT-epoxy composites

### 3.1 Morphology of CNT Mats and CNT-epoxy composites

The CNT mat and CNT composite microstructures were imaged by scanning electron microscope (SEM), using a voltage of 5 kV and a spot size of 3 µm. A field emission transmission electron microscope (FEGTEM), fitted with energy-dispersive X-ray spectroscopy (EDX) was used to characterise the bundle microstructure of the CNT mat and CNT-epoxy composites. Samples of the as-received CNT mat were prepared for TEM analysis by depositing a small quantity of aerogel layers from the delaminated mat onto a copper mesh, with the aid of acetone solvent. Composites were also prepared for TEM analysis by using a modified epoxy resin containing uniformly distributed silicon side-groups[4], so that the distribution of epoxy in the composite could be identified by elemental EDX mapping of the silicon group distribution. A focused ion beam (FIB)[5] was used to mill samples of direct-spun CNT mat and CNT-epoxy composites in their through-thickness direction in order to reveal the bundle microstructure in the cross-sectional view. A beam current of 2.8 nA was used for refined milling prior to imaging. For TEM analysis, samples were also prepared by FIB cutting, using a lift-out process as described elsewhere [72]. This method produced samples of suitable thickness between 100 nm and 150 nm.

---

[4] SILIKOFTAL® ED, Evonik, Tego House, Chippenham Dr, Kingston, Milton Keynes, MK10 0AF, UK
[5] SEM/FIB Workstation, Helios Nanolab DualBeam 600. Thermo Fisher Scientific, 168 Third Avenue, Waltham, MA, USA 02451.

Plan views of CNT mat microstructure and CNT composite microstructure with selected values of epoxy content are presented in Figure 6(a-c), and images of their cross-section in the through-thickness direction are shown in Figure 6(d-f). The CNT mat microstructure, as displayed in Figure 6(a) and 6(d), consists of a random network of CNT bundles, with branching of CNTs from bundle to bundle. The bundle cross sections are typically circular. In composites of low epoxy content, the epoxy uniformly coats the CNT bundle network, see Figure 6(b) and 6(e); some of the CNT bundles appear flattened in cross-sectional view compared to the dry state. As the epoxy content increases, it progressively fills the air space between CNT bundles until close to fully dense, see Figure 6(c) and 6(f). SEM and TEM images of the CNT-epoxy composite cross-section reveal an almost uniform volume fraction $f_e$ of epoxy in the through-thickness direction.

A plan view image of a CNT bundle in the as-received direct-spun mat from TEM is shown in Figure 6(g), revealing the crystalline, close-packed CNT bundle microstructure. A complementary, transverse image of a CNT bundle cross section in the direct-spun mat/silicone epoxy resin composite, produced from a cross-sectional cut of the composite in the through-thickness direction, is given in Figure 6(h). The CNTs are predominantly multi-walled, and the individual CNTs within a bundle remain close-packed after epoxy infiltration. Elemental mapping of this bundle cross-sectional view is shown in Figure 6(i), revealing the distribution of silicon-tagged epoxy within the microstructure. The elemental mapping shows that the epoxy resin surrounds the bundle, and wets the bundle surface, but does not penetrate the gap between neighbouring CNTs of each bundle.

## 3.2 Uniaxial tensile tests

Uniaxial tensile tests were performed on the CNT mat, CNT-epoxy composites, and the epoxy matrix, using a screw-driven test machine[6] at a strain rate of $\dot{\epsilon} = 10^{-4}\,\text{s}^{-1}$. The ends of the specimens were placed in wedge grips and were cushioned by paper end tabs, see Figure 7(a). The axial nominal strain in the gauge section was measured by the axial displacement of two dot stickers. The dots were of diameter 0.5 mm and of spacing 30 mm, and were adhered to the sample prior to testing. The relative displacement of the dots was tracked at a frequency of 1 Hz by a suitable camera system[7]. For the as-received direct-spun CNT mat, the yield stress is inferred from the stress-strain response via a bilinear fit, whilst for the CNT-epoxy composites, it is determined by finding the intersection of the stress-strain curve with a line drawn parallel to the initial, linear curve but offset by 0.2% in strain. The dimensions of the cast epoxy samples for uniaxial tensile testing are given in Figure 7(b); the tensile strain was again measured with the optical extensometer as described above.

The tensile uniaxial stress-strain response of CNT mat, CNT-epoxy composites and epoxy are plotted in Figure 8(a)-(c). In order to determine the degree of in-plane anisotropy, tensile tests were conducted on samples of dry mat (1) and composite (9), oriented at 0°, 45° and 90° to the principal material orientation, see Figure 8(a). Both materials exhibit moderate in-plane anisotropy. The ductility of the composite is inferior to that of the as-received mat, but the strength and stiffness are both enhanced. The uniaxial tensile responses of CNT mat-epoxy composite samples oriented at 0° to the principal material direction, and consolidated under atmospheric vacuum, are compared with those of the as-received CNT mat and cured epoxy dogbone samples in Figure 8(b). Note that the stress-strain responses of the composites are much higher than those of the epoxy and direct-spun mat. The mechanical behaviour of the epoxy dogbone samples is almost insensitive to the initial concentration of epoxy in acetone, prior to evaporation of the acetone, see samples (10) and (11) in Figure 8(b). The stress-strain response of samples oriented at 0° to the principal material direction and subjected to a 10 MPa pressure prior to cure are compared with those that have been consolidated under vacuum in Figure 8(c). Consolidation with a 10 MPa through-thickness pressure results in superior mechanical properties for both the direct-spun mat and composites; these composites exhibit the highest strength of 410 MPa and stiffness of almost 30 GPa in the present study.

---

[6] Instron Ltd, Coronation Road, High Wycombe, Buckinghamshire, HP12 3SY, UK. A load cell of maximum capacity 500 N was used for all tests.
[7] GOM UK Ltd, 14 Siskin Parkway East, The Cobalt Centre, Coventry, CV3 4PE, UK.

The modulus, yield strength and ultimate strength of CNT mat-epoxy composites, direct-spun mat, and epoxy are plotted against the measured sample density in Figure 8(d), (e) and (f), respectively. The modulus and strength are dependent on the volume fraction of both CNTs and epoxy in a non-linear manner. The greatest measured moduli and yield strengths of CNT-epoxy composites lie approximately an order of magnitude above the corresponding values for the unreinforced, as-received CNT mat, or the epoxy measured in bulk form. Also, the high pressure cure cycle leads to an increase in modulus and strength by a factor of about 2 for the same density.

**3.3 In-plane electrical properties, unloading and piezoresistive behaviour**

The in-plane electrical conductivity of the CNT mat and of the CNT-epoxy composites were measured by a four-point probe method prior to mechanical tensile testing; the sample dimensions and experimental setup are described in Figure 7(c). Contacts of adequately low resistance for 4-point testing were made by laying the samples on copper contacts. The presence of breather layer material on each side of the composites during cure meant that excess resin was removed by the applied vacuum and reliable electrical contacts were achieved. The only instance where the electrical contact did require improvement was for composite (9); in this case, mild sanding of the surface was sufficient to ensure a satisfactory electrical contact. To verify that the electrical current density was constant over the sample cross-section, measurements of the potential drop along the gauge length were taken both on the side of the sample on which the current probes were placed, and on the opposite side. It was noted that current flow was uniform throughout the thickness.

The in-plane electrical conductivity measured from samples of direct-spun CNT mat (1) oriented at 0°, 45° and 90° to the principal material orientation, and of CNT-epoxy composite (9), are given in the bar chart of Figure 10(a); the electrical conductivity of both the direct-spun mat and composite exhibit similar levels of anisotropy. The electrical conductivity of direct-spun CNT mats and composites measured along the principal material orientation is plotted against the CNT volume fraction $f_B$ in Figure 10(b): the electrical conductivity $K$ scales linearly with CNT volume fraction $f_B$.

A limited number of tensile tests were also conducted with simultaneous measurement of the electrical resistance within the central portion of the sample gauge length, to compare the in-plane piezoresistive behaviour and post-yield mechanical response of the CNT mat (1) and CNT-epoxy composite (9); the results are given in Figure 10(c). Current and voltage probes were attached to the sample with electrically conductive silver paint, and the resistance was measured at a frequency of 1 Hz. A constant current of 25 mA was supplied to the sample throughout the test, and partial unloading was conducted at regular intervals. The measured sample resistance $R$ is normalised in Figure 10(c) by its value $R_0$ at the start of the test. The unloading modulus of the samples $E_u$ as defined by the gradient of the stress-strain response at the onset of unloading, are plotted in Figure 10(d) as a function of axial strain. The unloading modulus of the CNT mat increases with increasing strain, and this is due to the progressive alignment of CNT bundles with the loading direction. In contrast, the unloading modulus of the composite decreases with increasing strain. This is suggestive of microstructural damage, presumably in the form of cracking of the epoxy matrix [2].

## 4. A micromechanical model for modulus and yield strength

A micromechanical model is now developed, based on the characterisation of the underlying material microstructure, to understand the origin of the stiffening and strengthening due to the epoxy matrix within the CNT mat.

**4.1 An idealisation of CNT mat-epoxy composite microstructure**

The unreinforced CNT mat microstructure consists of an interlinked network of nanotube bundles. The bundles are connected by the branching of CNTs from one bundle to the next. This random, interconnected, dry bundle network has a nodal connectivity of between 3 and 4, and earlier in-situ experiments [57] reveal that it deforms in a foam-like manner, predominantly due to bending and shearing of the CNT bundles, rather than by their axial stretch. For an estimation of strength and stiffness, this microstructure motivates the use of a periodic, planar hexagonal honeycomb network, represented by a repeating honeycomb unit cell of interlinked CNT bundle struts and epoxy, as illustrated in Figure 9(a). This unit cell was used before as an idealisation for dry CNT mat, and provided a useful estimate for the dry CNT mat modulus and yield strength [57]. Microscopy of the composites reveals that the outer surfaces of the CNT bundles are coated with epoxy, and that the epoxy does not infiltrate the bundles, see Figure 6(h) and (i). An increase in volume fraction of epoxy leads to progressive filling of the pores between the bundles. Here, we vary the volume fractions of CNT bundles, epoxy and air in the unit cell according to their values recorded in experiment, and assume that the epoxy extends from the CNT bundle struts into the centre of the pores, with thickness equivalent to that of the bundles in the out-of-plane direction, as shown in Figure 9(b). Finite element calculations were conducted to model the stress-strain response of this honeycomb unit cell in plane stress, using the commercially available finite element package ABAQUS Standard (version 6.14). The simulation set up and boundary conditions are illustrated in Figure 9(c); the roller boundary conditions were defined such that there are vanishing components of stress in the $x_2$ direction, and to achieve symmetry and deformation consistent with that of the periodic honeycomb network. A perfect-bond between the CNT bundles and epoxy is assumed. The degree of anisotropy of response of the unit cell is controlled by suitable choice of the initial value of the angle $\omega$, which is a measure of the alignment of the CNT bundle microstructure with the principal material direction. The thickness of the CNT bundle struts and epoxy layer are listed in Table S2 of the supplementary information for the simulations, alongside illustrations of the simulated unit cell over the compositional range of CNT-epoxy composites and direct-spun mat in Figure S2.

**4.2 Constitutive model for CNT bundles and matrix**

Covalent bonding within the CNT walls endows them with high axial strength and stiffness. In contrast, the bonds between adjacent CNT tubes are comparatively weak, and hence the longitudinal shear modulus and shear strength of CNT bundles is much below their axial modulus and strength in tension [3,73,74]. Here, we recognise that CNT bundles are transversely isotropic with respect to their longitudinal axis, and treat the CNT bundle as an anisotropic, homogeneous continuum. We define the constitutive relationships as follows.

Consider first the elastic response of CNT bundles. The elastic strain $\varepsilon_{ij}^e$ is related to the stress $\sigma_{ij}$ by:

$$\begin{Bmatrix} \varepsilon_{11}^e \\ \varepsilon_{22}^e \\ \varepsilon_{33}^e \\ \gamma_{23}^e \\ \gamma_{13}^e \\ \gamma_{12}^e \end{Bmatrix} = \begin{bmatrix} 1/E_{11}^B & -\nu_{21}/E_{22}^B & -\nu_{31}/E_{33}^B & 0 & 0 & 0 \\ -\nu_{12}/E_{11}^B & 1/E_{22}^B & -\nu_{32}/E_{33}^B & 0 & 0 & 0 \\ -\nu_{13}/E_{11}^B & -\nu_{23}/E_{22}^B & 1/E_{33}^B & 0 & 0 & 0 \\ 0 & 0 & 0 & 1/G_{23}^B & 0 & 0 \\ 0 & 0 & 0 & 0 & 1/G_{13}^B & 0 \\ 0 & 0 & 0 & 0 & 0 & 1/G_{12}^B \end{bmatrix} \begin{Bmatrix} \sigma_{11} \\ \sigma_{22} \\ \sigma_{33} \\ \tau_{23} \\ \tau_{13} \\ \tau_{12} \end{Bmatrix} \quad (2)$$

We proceed to obtain estimates for the elastic constants in (2). Further details of our methods and constants used in calculation are provided in Section 2 of the supplementary information. The modulus of CNT walls measured in axial tension, and based upon an assumed interlayer spacing of 0.335 nm, is taken to be 1 TPa [3,75,76]. We estimate the bundle axial modulus by assuming all CNT walls within the bundle cross-section are subjected to a uniform tensile strain. The longitudinal CNT bundle modulus $E_{11}^B$ is related to the wall modulus $E_w$ and wall density $\rho_w$ by $E_{11}^B = E_w(\rho_B/\rho_w)$ [77]. Substitution of the measured values of measured bundle density $\rho_B$ and wall density $\rho_w$ = 2300 kg/m³ implies that $E_{11}^B$ = 680 GPa. Suggested values within the literature [78,79] for the transverse modulus $E_{22}^B = E_{33}^B$ for bundles of single-walled CNTs range from 40 GPa to 78 GPa. Here, we assume that $E_{22}^B = E_{33}^B$ = 50 GPa, and take the Poisson ratio to be $\nu_{12} = \nu_{13} = 0.3$. The low values of bundle shear moduli $G_{12}^B$, $G_{13}^B$ and $G_{23}^B$ all result from the weak interfacial bonding between CNTs, and we assume that they are all equal. By calibration of the predicted macroscopic stiffness of the CNT honeycomb with the measured modulus of the dry unreinforced CNT mat upon unloading, we already deduced in a previous study [57] that $G_{12}^B$ = 9.5 GPa, and this value is again used herein.

Now consider the strength and post-yield behaviour of a CNT bundle. Hill's anisotropic yield criterion [80] is used here to represent the post-yield behaviour of the CNT bundles. The total strain rate $\dot{\varepsilon}_{ij}$ is the sum of the elastic strain rate $\dot{\varepsilon}_{ij}^e$ and plastic strain rate $\dot{\varepsilon}_{ij}^p$,

$$\dot{\varepsilon}_{ij} = \dot{\varepsilon}_{ij}^e + \dot{\varepsilon}_{ij}^p. \tag{3}$$

The plastic strain rate is defined by the associated flow rule,

$$\dot{\varepsilon}_{ij}^p = \dot{\lambda}\frac{\partial \Phi}{\partial \sigma_{ij}}, \tag{4}$$

in terms of a plastic multiplier $\dot{\lambda}$, and the Hill potential $\Phi$, as defined by:

$$2\Phi = F(\sigma_{22} - \sigma_{33})^2 + G(\sigma_{33} - \sigma_{11})^2 + H(\sigma_{11} - \sigma_{22})^2 + 2L\tau_{23}^2 + 2M\tau_{31}^2 + 2N\tau_{12}^2 \tag{5}$$

The constants $F$, $G$ and $H$ are directly related to the tensile yield strength of the CNT bundle in uniaxial tension, $\sigma_{11}^B$, $\sigma_{22}^B$ and $\sigma_{33}^B$, such that

$$G + H = \frac{1}{(\sigma_{11}^B)^2}; \quad F + H = \frac{1}{(\sigma_{22}^B)^2}; \quad G + F = \frac{1}{(\sigma_{33}^B)^2}. \tag{6}$$

The constants $L$, $M$ and $N$ follow from the shear yield strengths, where

$$L = \frac{1}{2(\tau_{23}^B)^2}; \quad M = \frac{1}{2(\tau_{31}^B)^2}; \quad N = \frac{1}{2(\tau_{12}^B)^2}. \tag{7}$$

Experiments on individual CNT bundles reveal that the tensile wall fracture strength $\sigma_w$ is between 5.5 GPa and 25 GPa [81]. Here, we estimate the bundle axial strength by assuming that it, like the axial bundle modulus, scales with the CNT wall strength and bundle density, such that $\sigma_{11}^B = \sigma_w(\rho_B/\rho_w)$. Upon taking $\sigma_w = 5.5$ GPa [81], we estimate the axial bundle fracture strength to be 3.7 GPa. The longitudinal and transverse shear yield strengths, $\tau_{12}^B$, $\tau_{13}^B$ and $\tau_{23}^B$, and the remaining transverse normal yield strengths $\sigma_{22}^B$ and $\sigma_{33}^B$, are set equal to $\tau_y^B$, which is the macroscopic as-measured yield strength. A hardening modulus of value $10^{-4}E_{11}^B$ for all stress components was employed post-yield to ensure converged results. The epoxy matrix is treated as an isotropic elastic, perfectly plastic solid that satisfies $J_2$ flow theory. A summary of all material properties used in the finite element simulations are listed in Table 2.

## 4.3 Calibration of the unit cell model

The honeycomb model was calibrated via the following steps.

Step (I): the elastic response of the dry honeycomb (absent epoxy) was determined for uniaxial loading in the $x_1$ and $x_2$ directions, in order to obtain $E_{11}/E_{22}$ for $30° \leq \omega \leq 50°$. The sensitivity of $E_{11}/E_{22}$ to the initial value of $\omega$ is shown in Figure 11(a), and the measured ratio $E_{11}/E_{22} = 6.4$ implies that $\omega = 45°$.

Step (II): The uniaxial yield strength of the honeycomb unit cell was predicted for compositions (1) and (9), again for loading in the $x_1$ and $x_2$ directions. The initial inclination of the unit cell wall $\omega$ was set to be 45° for all simulations, but the bundle shear strength $\tau_y^B$ was varied. By matching the predicted macroscopic yield strength $\sigma_{11}^y$ to the measured value of the CNT dry mat (1), we deduce $\tau_y^B$ = 80 MPa. Likewise, by matching the predicted value of $\sigma_{11}^y$ to the measured value for the CNT-epoxy composite (9) we deduce that $\tau_y^B = 250$ MPa.

## 4.4 Prediction of the calibrated model

It remains to compare the predictions of the honeycomb model with the measured uniaxial response of the CNT-epoxy composites. The measured and predicted stress-strain responses for tensile loading in the $x_1$ and $x_2$ directions are shown in Figure 11(b) for dry CNT mat (1) with $f_e = 0$, CNT-epoxy composite (4) with $f_e = 0.14$, and CNT-epoxy composite (9) with $f_e = 0.79$. Adequate agreement is evident including the degree of anisotropy in yield behaviour. We emphasise that the approach is an approximation for the detailed geometry of the network and so precise quantitative agreement is not to be expected.

We now compare finite element predictions for the modulus and yield strength in the principal material direction over the range of manufactured mat and composite compositions.

The predictions and experimental measurements of modulus and yield strength are plotted against bulk density in Figure 11(c) and (d) respectively. In broad terms, the predictions of macroscopic modulus and yield strength from the unit cell model capture the trend in the experiments over the compositional range, although some scatter is present in the experimental data.

The measured modulus $E_{11}$ and yield strength $\sigma_{11}^{YS}$ of the CNT-epoxy composites are plotted against epoxy volume fraction $f_e$ in Figure 12(a) and (b), respectively. For composites with $0.17 \leq f_B \leq 0.2$, both $E_{11}$ and $\sigma_{11}^{YS}$ increase by about a factor of about 5 as the epoxy content is increased. The presence of epoxy within the pores of the CNT bundle network restricts the foam-like deformation observed for CNT mats absent epoxy [57], and this enhances the modulus and strength. For the range of epoxy content $0.3 \leq f_e \leq 0.4$, it is also clear that an increase in CNT bundle volume fraction from 0.25 to 0.35 increases $E_{11}$ and $\sigma_{11}^{YS}$.

Our unit cell model suggests a significant increase in the longitudinal shear strength of the CNT bundles $\tau_y^B$ from 80 MPa to 250 MPa upon the coating of CNT bundles with epoxy, and the source of this increase in bundle shear strength is now discussed. In composites of direct-spun CNT mats and epoxy, the epoxy bonds strongly to the surfaces of the CNT bundle reinforcement [42,82,83], and a thin interfacial layer of enhanced strength forms on the surfaces of the bundles, as sketched in Figure 12(c). It is known from fractography [33] that the layer of epoxy surrounding the bundles has a yield strength much above that of the epoxy matrix which fills the voids within the bundle network: Images of the fracture surface of direct-spun CNT mat-epoxy composites reveal that CNT bundles protruding from the fracture surface are coated with this sheath of epoxy. Pull-out tests upon individual CNTs embedded in epoxy reveal that the strength of the epoxy layer which coats CNTs can be over 350 MPa [84,85], even reaching 630 MPa [86]. The epoxy adheres adjacent CNTs within the outer layer of the bundle, and thereby increases the bundle longitudinal shear strength. Note that this effect is modelled in the 2D context by increasing the shear strength of the CNT bundles, $\tau_y^B$.

## 5. Concluding discussion

The modulus and strength of CNT-polymer composites as reported in the literature vary over several orders of magnitude, and are sensitive to both composition and microstructure. However, all reported data for the modulus, strength, electrical conductivity and thermal conductivity of CNT-polymer composites are much below those of individual CNTs.

In the present study, composites were manufactured by infiltration of direct-spun CNT mats with solutions of epoxy and acetone. Following curing, the composition of the manufactured composites varied widely; the volume fraction of CNT bundles varied between 0.11 and 0.35, the volume fraction of epoxy from 0.01 to 0.79, and the porosity ranged from 0.10 to 0.82. The epoxy content in the cured composite increased with the concentration of epoxy in the infiltration solution; the application of a pressure $P = 10$ MPa in the through-thickness direction after infiltration was used to attain higher CNT volume fractions. The epoxy matrix does not penetrate the CNT bundles, but progressively fills the pores between the CNT bundles as the epoxy volume fraction increases.

The modulus, yield strength and ultimate strength of the CNT mat-epoxy composites are significantly above those of the dry CNT mat or cured epoxy. By suitable choice of composition, composites exhibit an ultimate strength of 410 MPa, and modulus of almost 30 GPa. Both values represent an increase in over an order of magnitude compared to the properties of the as-received CNT mat. The measured electrical conductivity of the mats and composites scale linearly with CNT volume fraction and is insensitive to the epoxy content.

The CNT-epoxy composite is idealised as a periodic honeycomb network in finite element simulations. The model is able to describe the degree of elastic and plastic anisotropy of the composite and the dependence of modulus and yield strength upon composition. By suitable correlation of the predicted and measured yield strengths of the composite, the inferred shear strength of a bundle is found to rise from 80 MPa in the absence of epoxy to 250 MPa when epoxy is present. These values are consistent with those reported in the literature. We deduce from comparison of measured composite properties against composition, and via simulation, that the properties of the composite are sensitive to the coating of CNT bundles with an interfacial layer of high strength epoxy, the epoxy volume fraction within the pores, and the CNT bundle volume fraction.

## Acknowledgements

W. Tan and J.C. Stallard contributed equally to this work. The authors would like to gratefully acknowledge the funding from the EPSRC project 'Advanced Nanotube Application and Manufacturing (ANAM) Initiative' under Grant No. EP/M015211/1, and for financial support from the ERC 'Multi-phase Lattice Materials' (MULTILAT) under Grant No. 669764. The authors would also like to acknowledge Tortech Nano Fibers Ltd for supplying CNT mat material.

# Figures

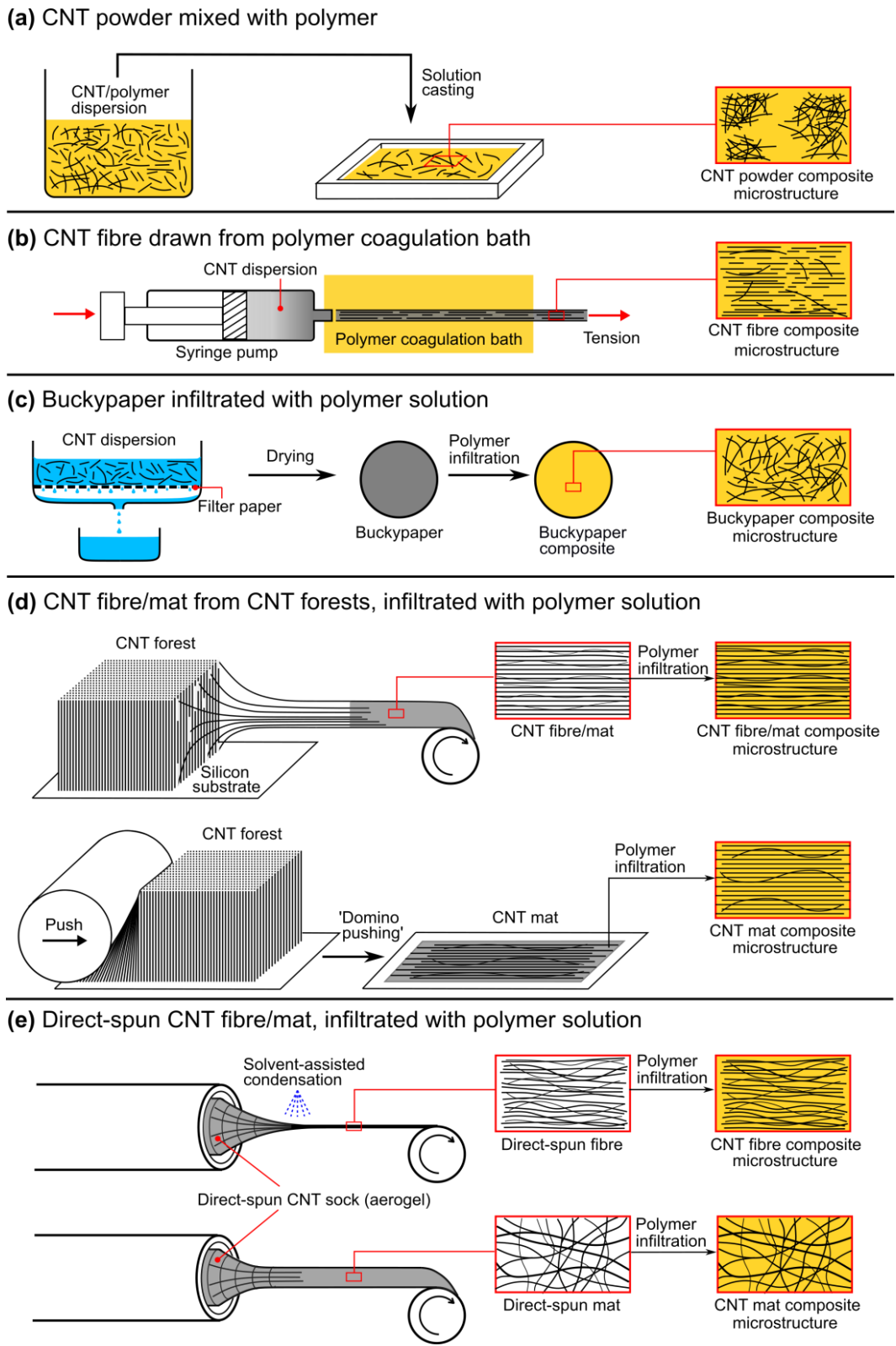

Figure 1: Classes of CNT polymer composites.

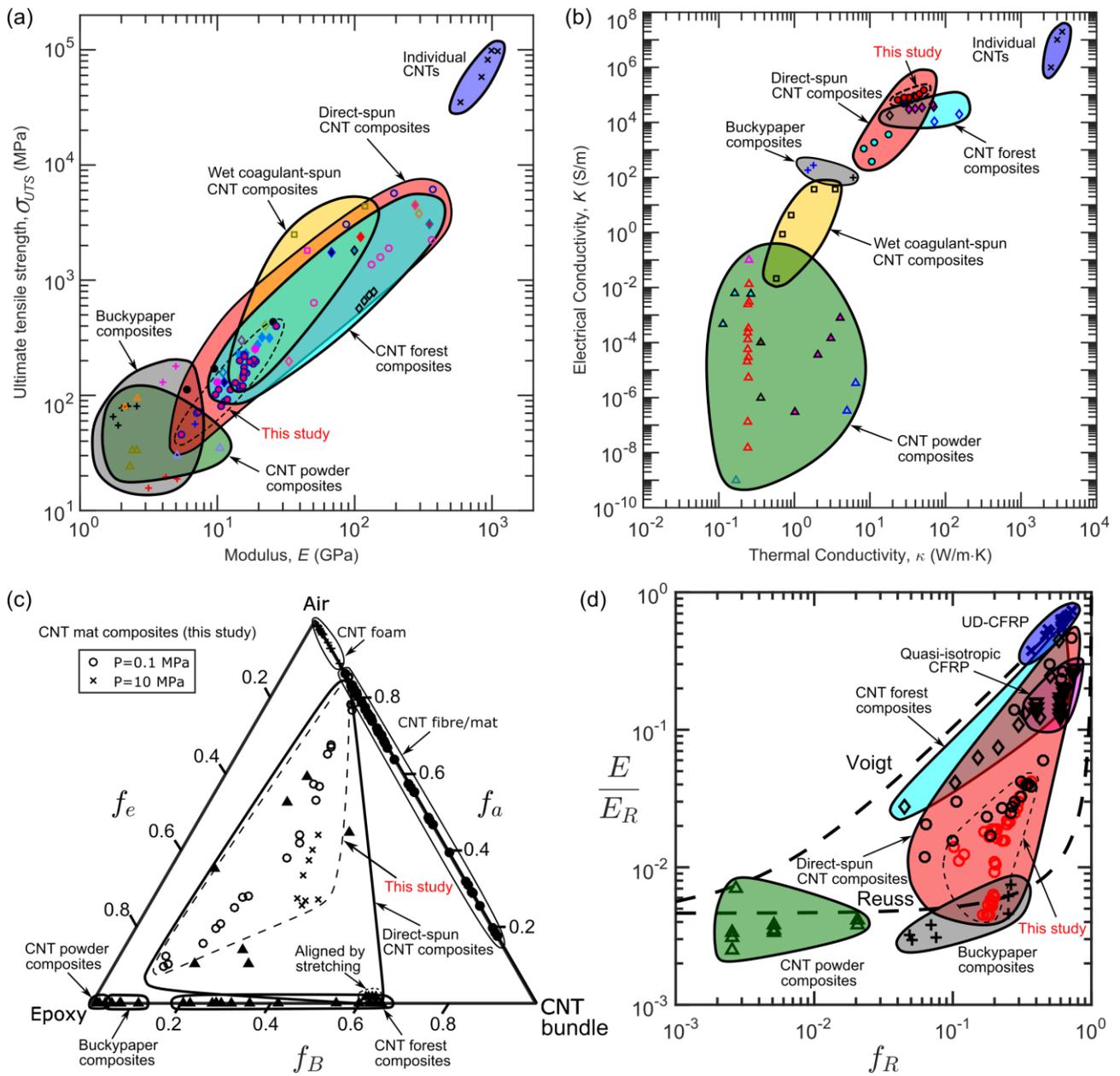

Figure 2: The mechanical properties (a) and electrical and thermal properties (b) of CNTs and CNT-polymer composites. (c) Composition diagram for CNT foams, CNT fibres/mats, and CNT-epoxy composites. (d) The moduli of CNT-epoxy and CFRP composites, normalised by the reinforcement modulus, is plotted against fibre volume fraction. Data are taken from [8-67] and from the present study.

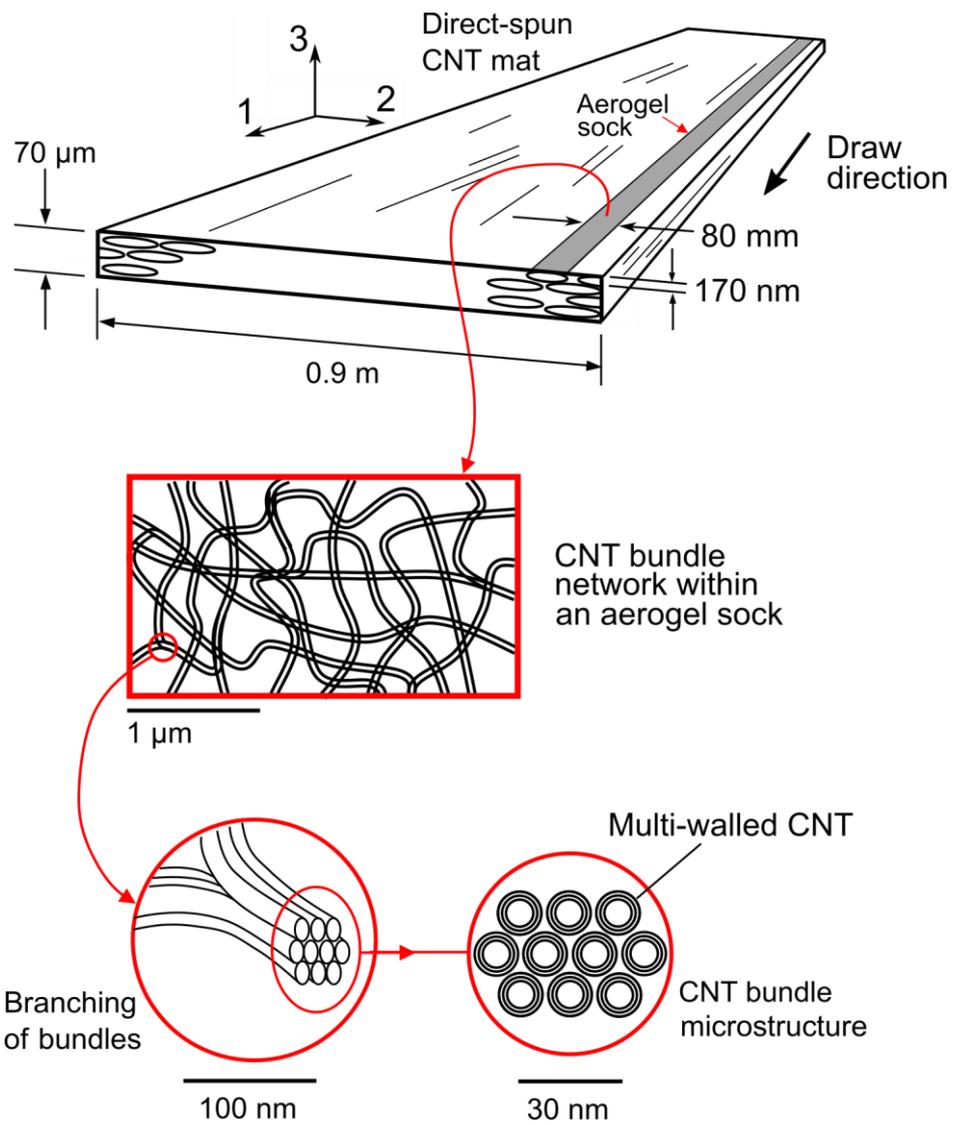

Figure 3: The hierarchical microstructure of direct-spun CNT mats.

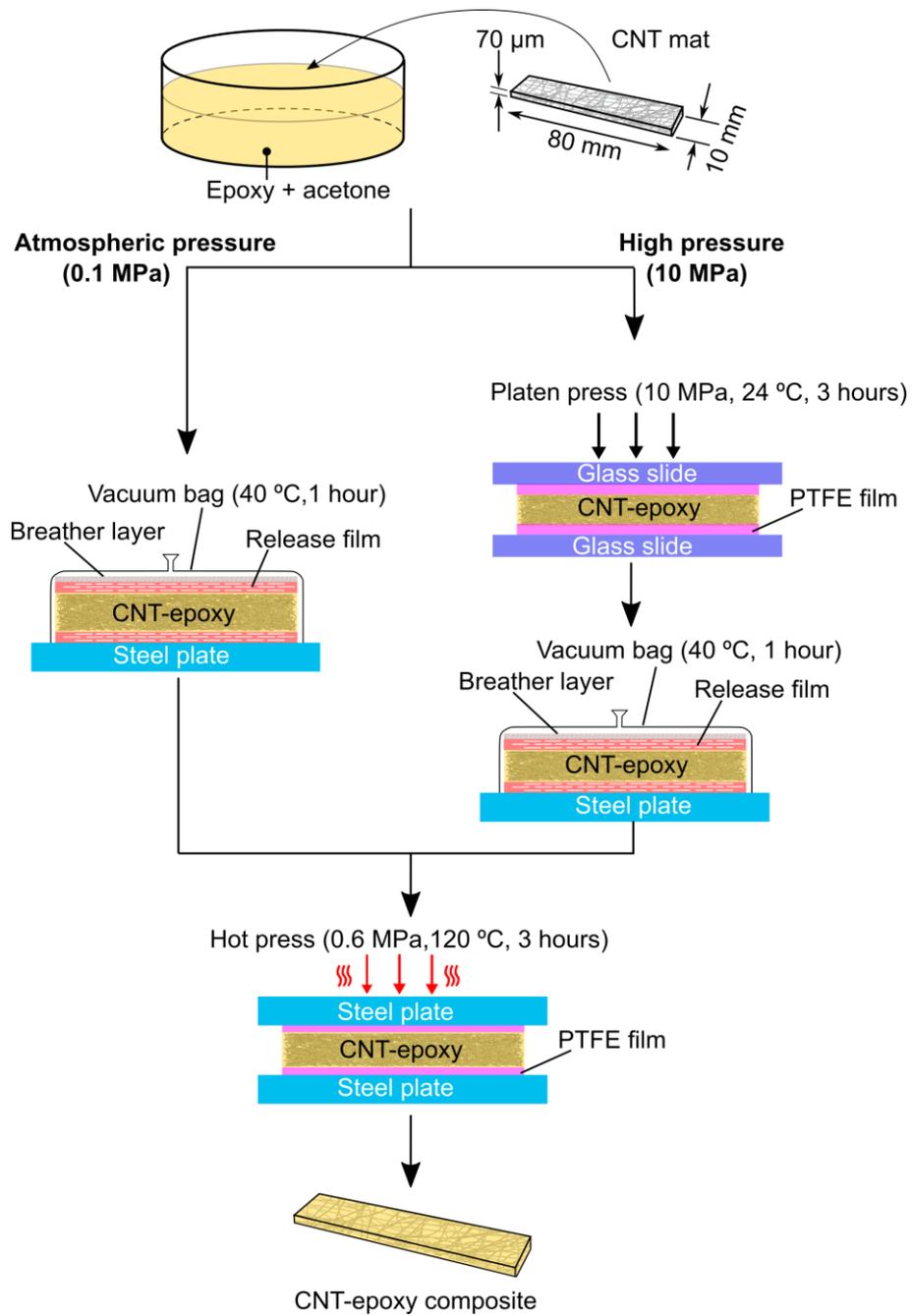

Figure 4: Methodology of CNT-epoxy composite manufacture.

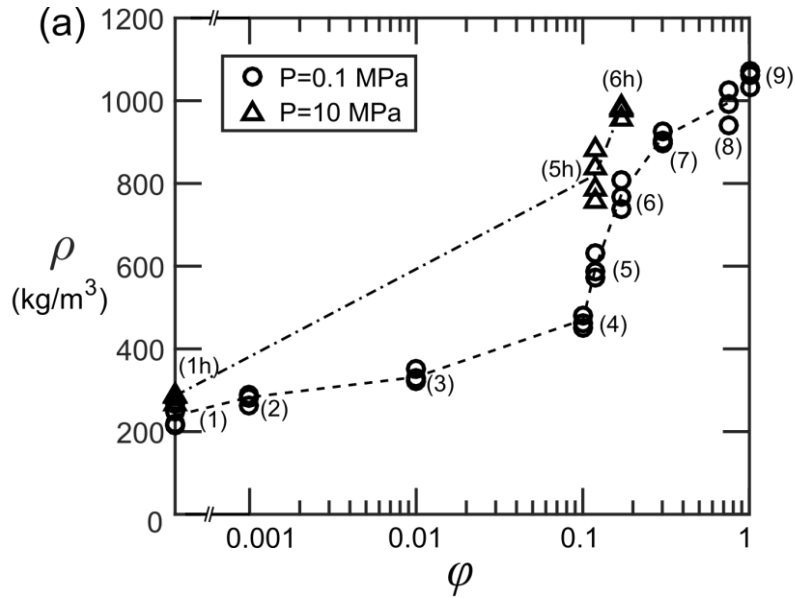
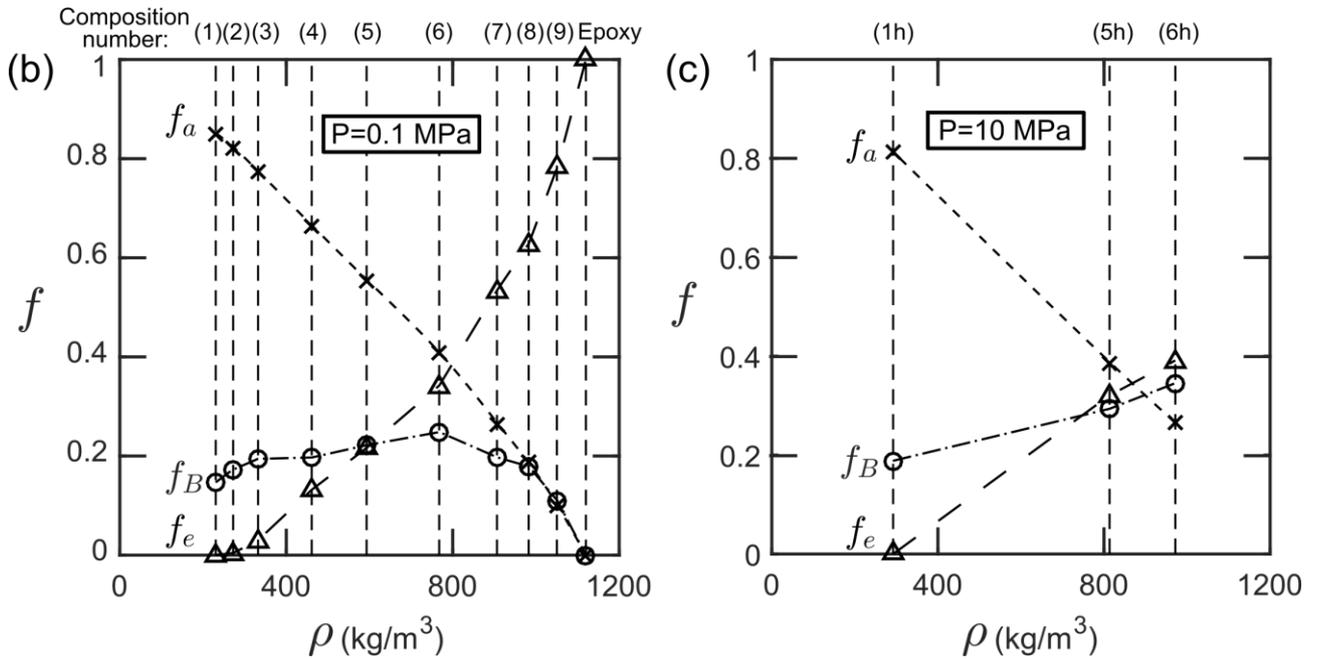

Figure 5: (a) Density $\rho$ of dry mat and cured composites, plotted against the concentration of epoxy by weight $\phi$ for the infiltration solution used in manufacture. The volume fractions $f$ of CNT bundles, epoxy and air in the cured composite are plotted against bulk composite density in (b) for samples manufactured with a consolidation pressure $P = 0.1$ MPa, and in (c) for samples subjected to a 10 MPa pressure prior to vacuum.

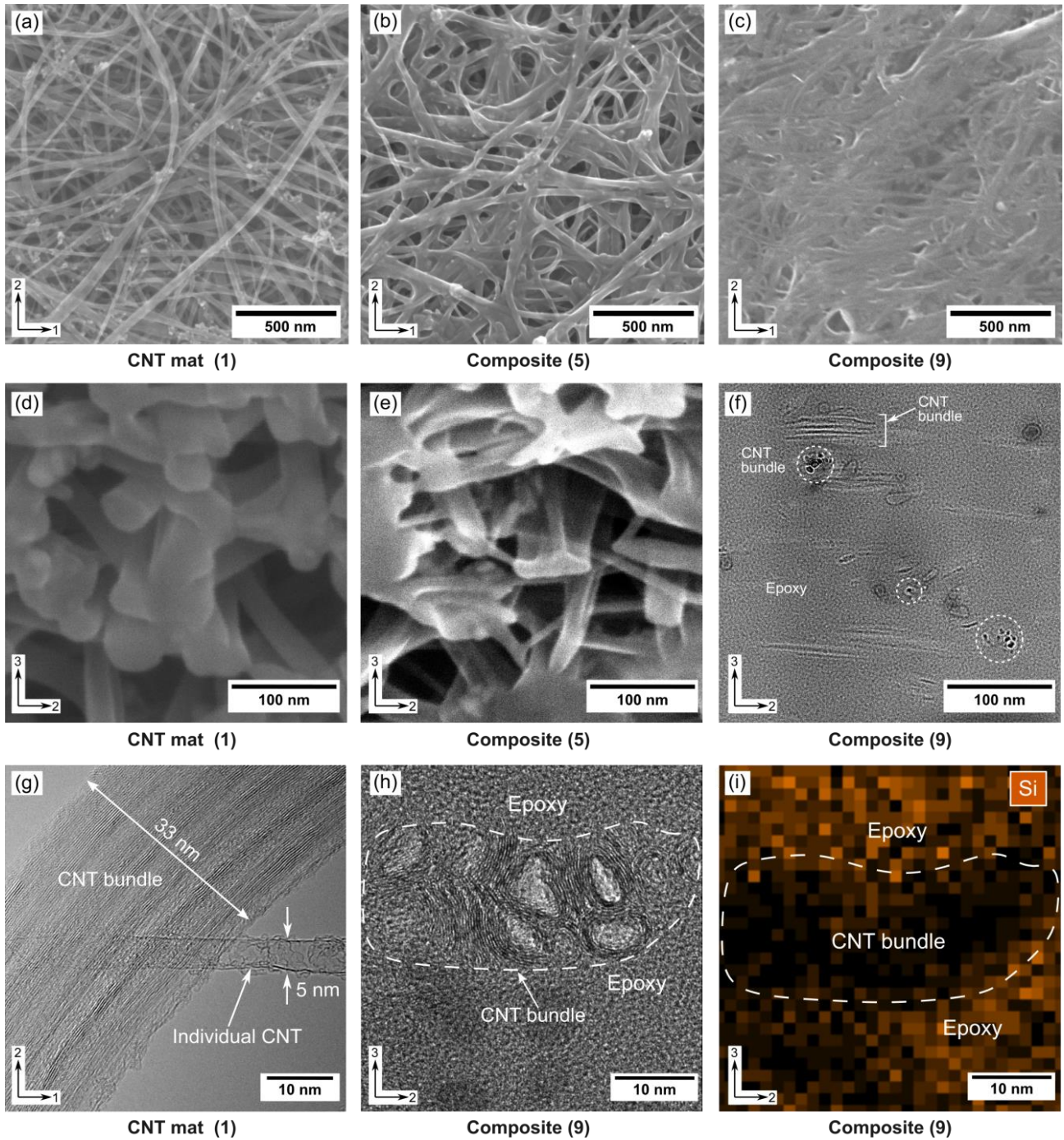

Figure 6: Plan-view of (a) CNT mat, (b) composite (5) and (c) composite (9). FIB-milled cross-sections for (d) dry mat, (e) composite (5) and (f) composite (9). (g) Plan view of a CNT bundle in dry mat; (h) FIB-milled cross-section of composite (9), with (i) distribution of Silicon-tagged epoxy. Images (a)-(e) taken in SEM, (f)-(i) in TEM.

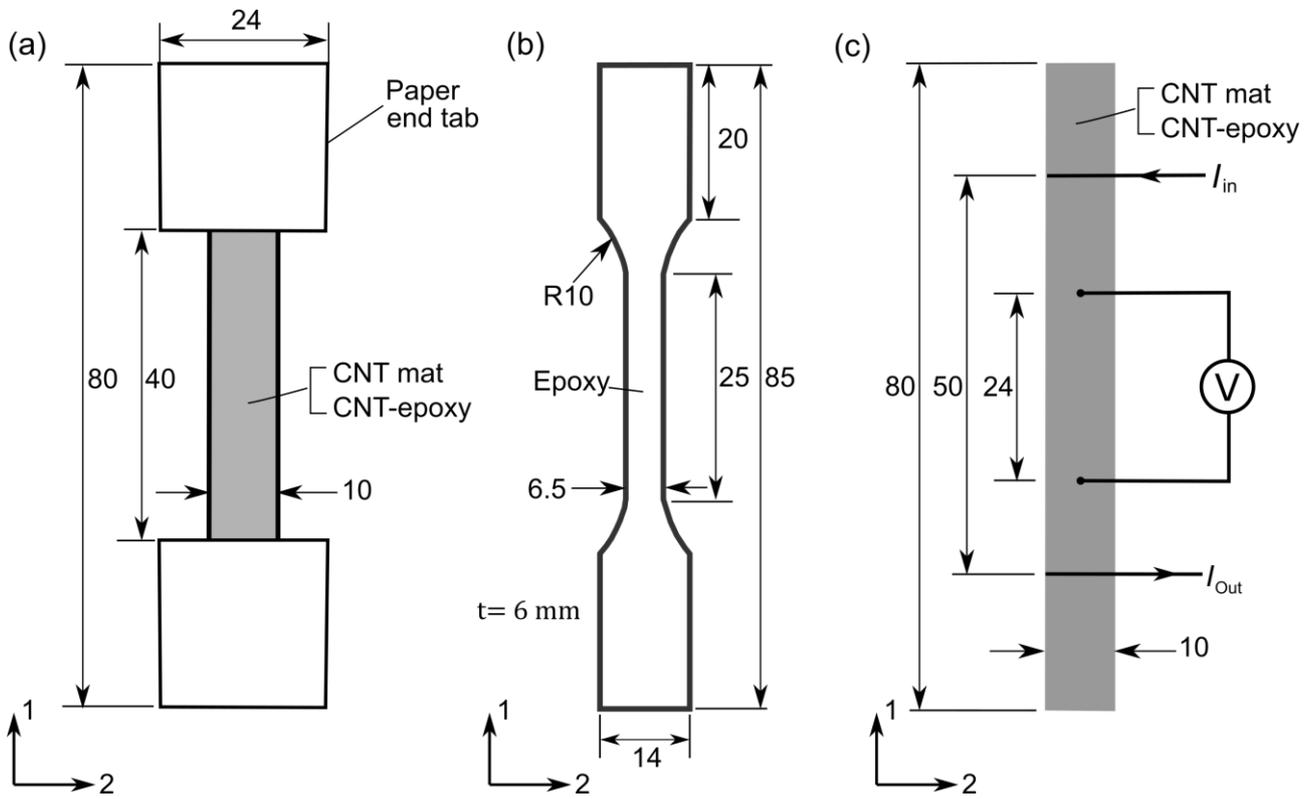

Figure 7: Specimen geometries, (a) for uniaxial tensile tests on CNT mat or CNT-epoxy composites, (b) epoxy dog-bone sample, with thickness 6 mm. (c) Four-point probe method used to measure the electrical conductivity. All dimensions are in mm.

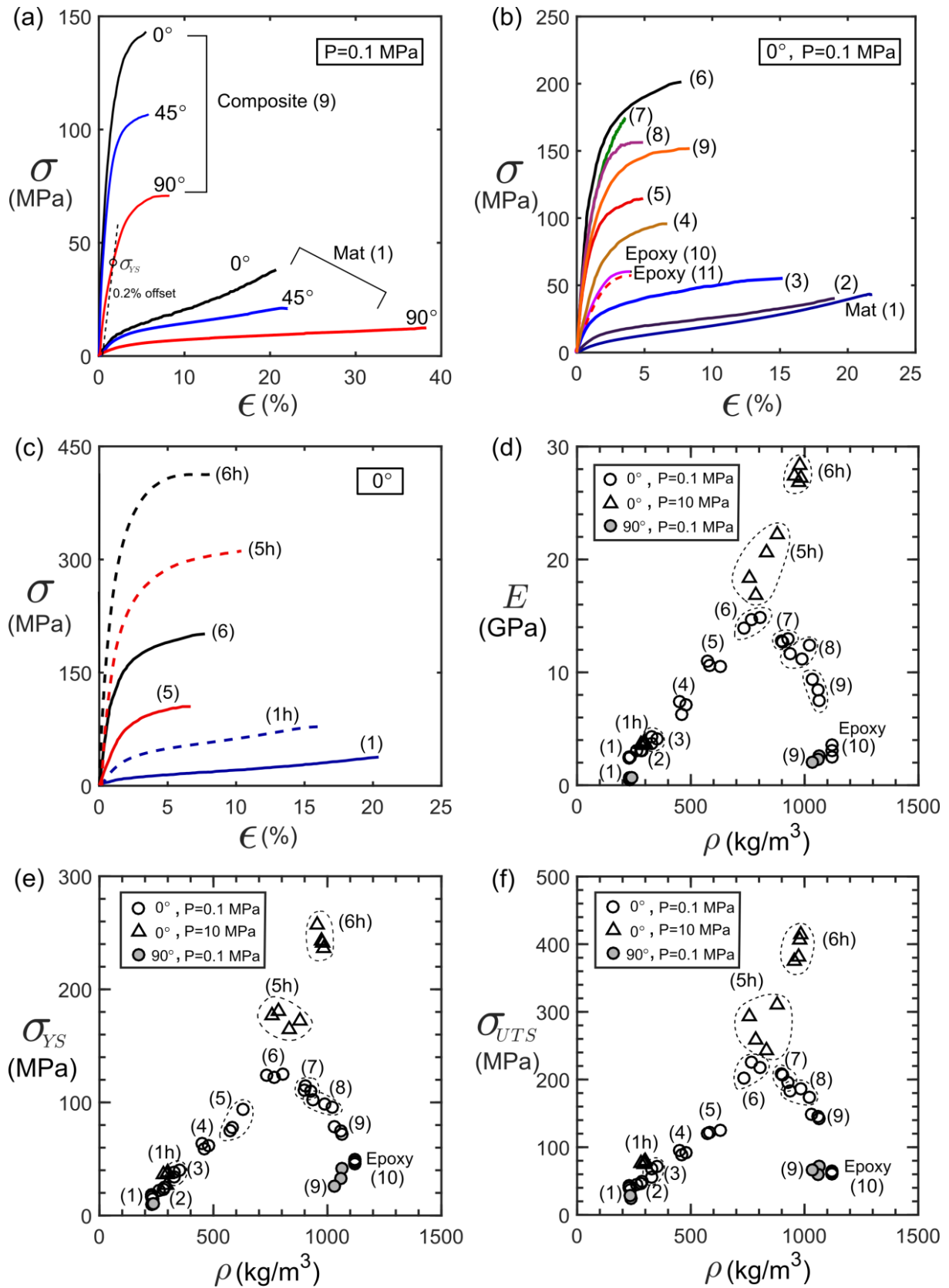

Figure 8: Stress versus strain response of composites and mat (a) in 3 orientations, (b) manufactured at $P = 0.1$ MPa. Materials manufactured with $P = 1$ MPa and $P = 10$ MPa are compared in (c). Dependence upon density $\rho$ of modulus $E$ in (d), yield strength $\sigma_{YS}$ in (e), and ultimate tensile strength $\sigma_{UTS}$ in (f) for the composites, direct-spun mat, and epoxy.

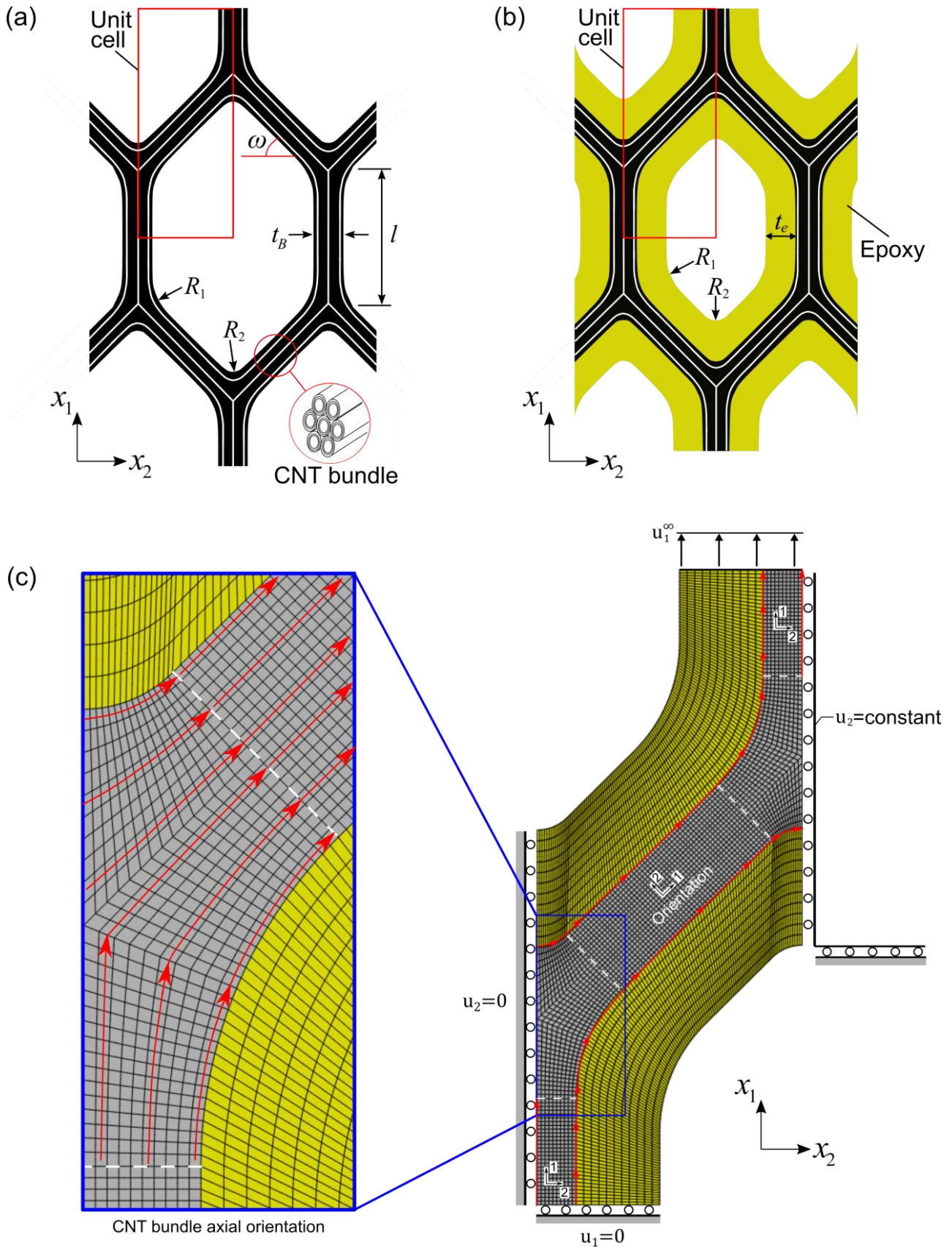

Figure 9: The planar honeycomb unit cell idealisation of (a) CNT mat and (b) CNT-epoxy composite microstructure with $\omega = 45°$. Details of the simulation setup and boundary conditions for the repeating unit cell analysed are shown in (c); note the variation of material orientation around the node.

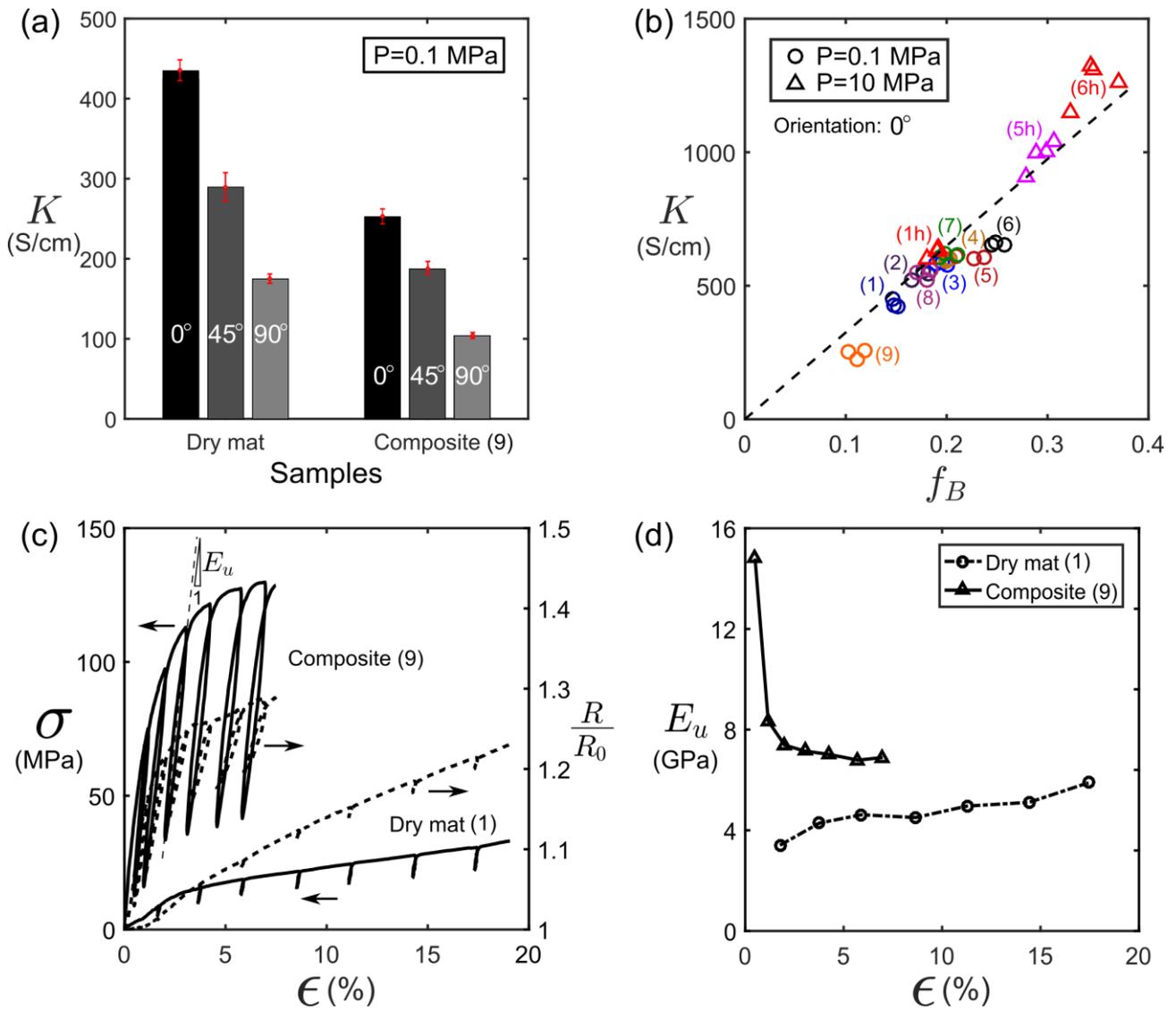

Figure 10: (a) The electrical conductivity of the CNT mat (1) and CNT-epoxy composite (9) in different directions. (b) Electrical conductivity in the principal material direction plotted against CNT volume fraction. (c) Piezoresistive and unloading response, showing evolution of sample resistance with strain for dry mat and composite. The moduli upon unload are plotted against the applied tensile strain in (d).

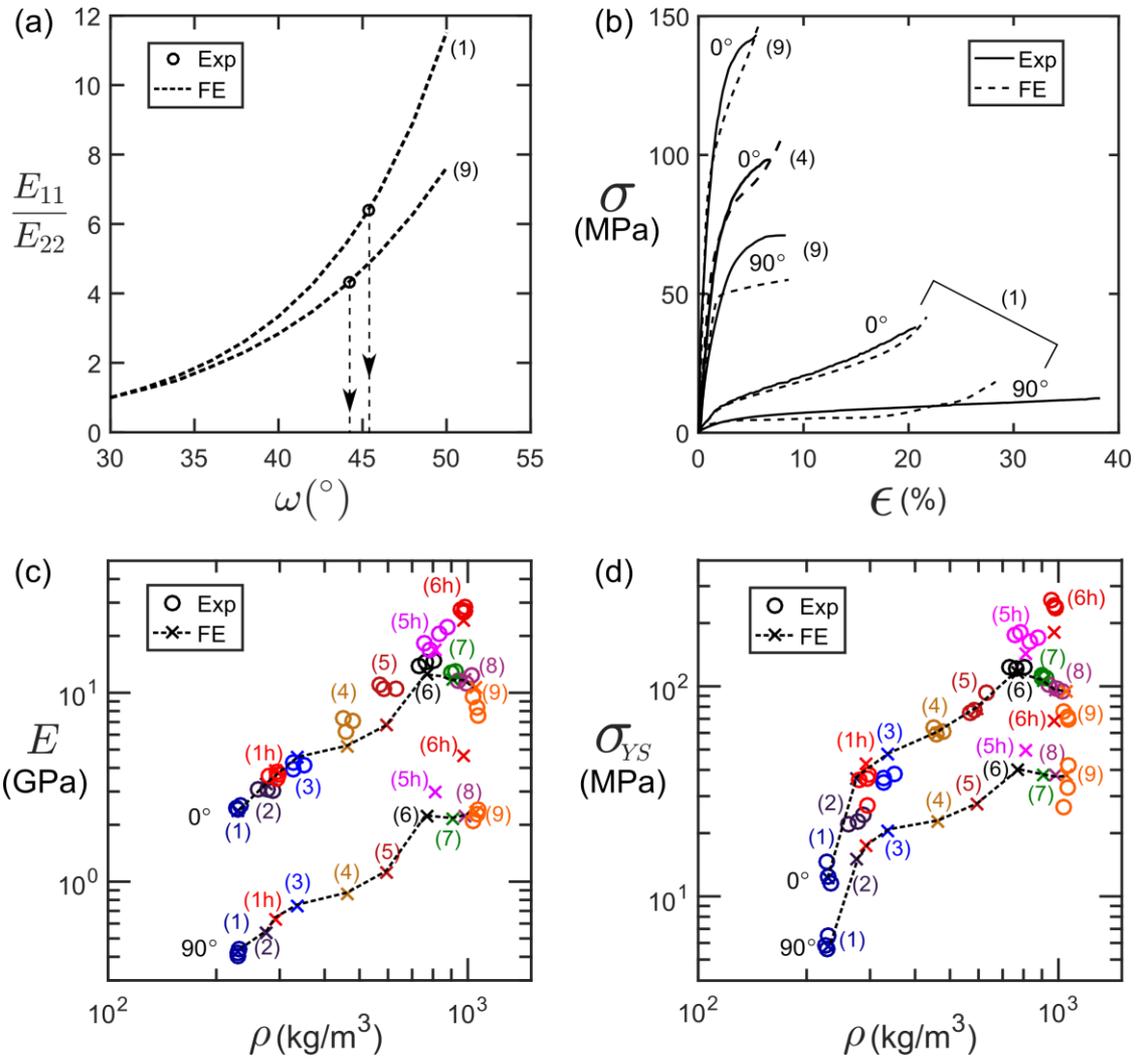

Figure 11: The ratios of the predicted principal and transverse moduli are plotted as a function of $\omega$ in (a), alongside the experimentally measured values. Measured and predicted uniaxial stress-strain responses are plotted in (b), and the modulus and yield strengths of manufactured CNT mat and composites in the principal material direction are compared with the unit cell predictions in (c) and (d) respectively.

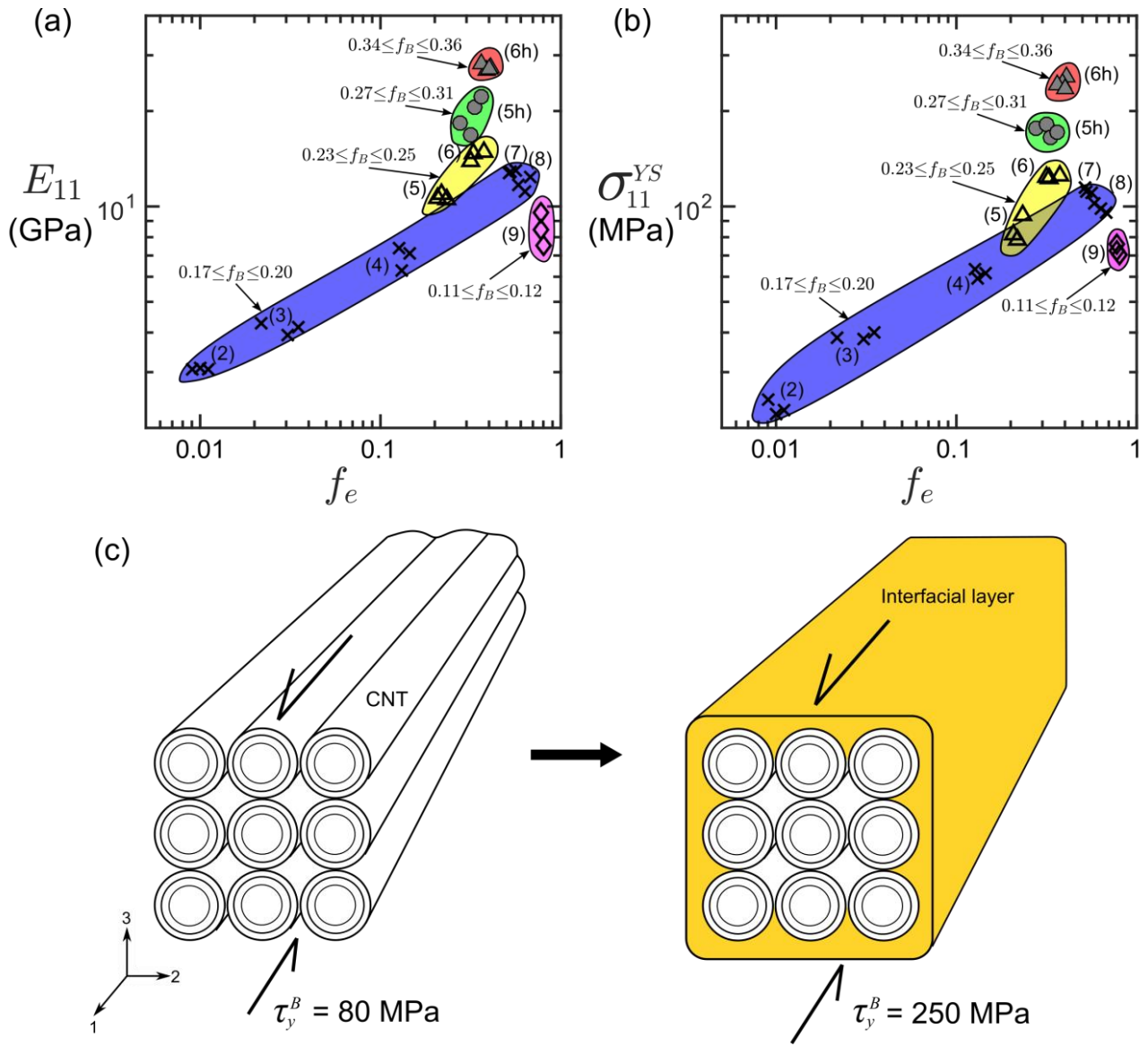

Figure 12: The measured modulus (a) and yield strength (b) for CNT-epoxy composites, as a function of composition. (c) CNT bundles are coated with an interfacial layer of enhanced yield strength upon infiltration.

# Tables

Table 1: Composition of direct-spun CNT mat and CNT-epoxy composites as a function of the process parameters

| Sample label | Epoxy concentration $\phi$ | Consolidation pressure $P$ (MPa) | Average density $\rho$ (kg/m³) | Average volume fraction | | |
|---|---|---|---|---|---|---|
| | | | | $f_B$ | $f_e$ | $f_a$ |
| (1)   | 0     | 0.1  | 234  | 0.15 | 0    | 0.85 |
| (1h)  | 0     | 10   | 296  | 0.19 | 0    | 0.81 |
| (2)   | 0.001 | 0.1  | 276  | 0.17 | 0.01 | 0.82 |
| (3)   | 0.01  | 0.1  | 330  | 0.19 | 0.03 | 0.78 |
| (4)   | 0.10  | 0.1  | 469  | 0.20 | 0.14 | 0.66 |
| (5)   | 0.12  | 0.1  | 605  | 0.23 | 0.22 | 0.55 |
| (5h)  | 0.12  | 10   | 811  | 0.29 | 0.32 | 0.39 |
| (6)   | 0.17  | 0.1  | 771  | 0.25 | 0.34 | 0.41 |
| (6h)  | 0.17  | 10   | 983  | 0.35 | 0.39 | 0.26 |
| (7)   | 0.30  | 0.1  | 906  | 0.20 | 0.53 | 0.27 |
| (8)   | 0.75  | 0.1  | 986  | 0.18 | 0.63 | 0.19 |
| (9)   | 1     | 0.1  | 1060 | 0.11 | 0.79 | 0.10 |
| (10)  | 1     | 0.1  | 1120 | 0    | 1    | 0    |
| (11)  | 0.50  | 0.1  | 1120 | 0    | 1    | 0    |

Table 2: Material constants used in the finite element analysis

| Material | Material constants |
|---|---|
| CNT bundle | $E_{11}^B = 680$ GPa, $E_{22}^B = E_{33}^B = 50$ GPa <br> $G_{12}^B = G_{23}^B = G_{13}^B = 9.5$ GPa, $\nu_{12} = \nu_{13} = 0.3$ <br> $\sigma_{11}^B = 3700$ MPa <br> $\sigma_{22}^B = \sigma_{33}^B = \tau_y^B = 80$ MPa (CNT bundle absent epoxy), 250 MPa (epoxy coated CNT bundle) <br> $\tau_{12}^B = \tau_{23}^B = \tau_{13}^B = \tau_y^B$ |
| Epoxy | $E^e = 3.0$ GPa, $\nu_e = 0.3$, $\sigma_y^e = 60$ MPa |

# Supplementary Information for

# THE MECHANICAL AND ELECTRICAL PROPERTIES OF DIRECT-SPUN CARBON NANOTUBE MAT-EPOXY COMPOSITES

Wei Tan, Joe C. Stallard, Fiona R. Smail, Adam M. Boies, Norman A. Fleck*

Address: Engineering Department, University of Cambridge, Trumpington Street, Cambridge, CB2 1PZ, UK.

*Corresponding author. Tel: +44 (0)1223 748240. Email: naf1@eng.cam.ac.uk (N.A. Fleck)

## 1. Axial elastic modulus of carbon fibres used in Figure 2

The value of reinforcement modulus $E_R$ used in normalisation of the longitudinal modulus of long fibre carbon fibre composites Figure 2(d) are given in Table S1.

Table S1: Axial elastic modulus of carbon fibres

| CFRP laminate | Ref. | Carbon fibre | Resin | $E_R$ (GPa) |
|---|---|---|---|---|
| Unidirectional | [1] | Modmor type I | LY558 epoxy | 410 |
| | [1] | Modmor type II | LY558 epoxy | 240 |
| | [1] | Rolls Royce fibres | LY558 epoxy | 200 |
| | [2] | Tenax HS45 fibre | LY556 epoxy | 240 |
| Quasi-isotropic | [3] | Sigmatex 450gsm | Gurit Prime 20LV | 212 |
| | [4,5] | AS4 | 8552, PEKK | 231 |
| | [6] | T300 | 5208,BP907,4901/(m)MDA | 230 |
| | [6] | T700 | 5208,BP907,4901/(m)MDA | 230 |

## 2. Thickness of mat and composites

The dependence of thickness $t$ upon composite density $\rho$ is plotted in Figure S1. The range and mean values are plotted for ten thickness measurements over the sample length.

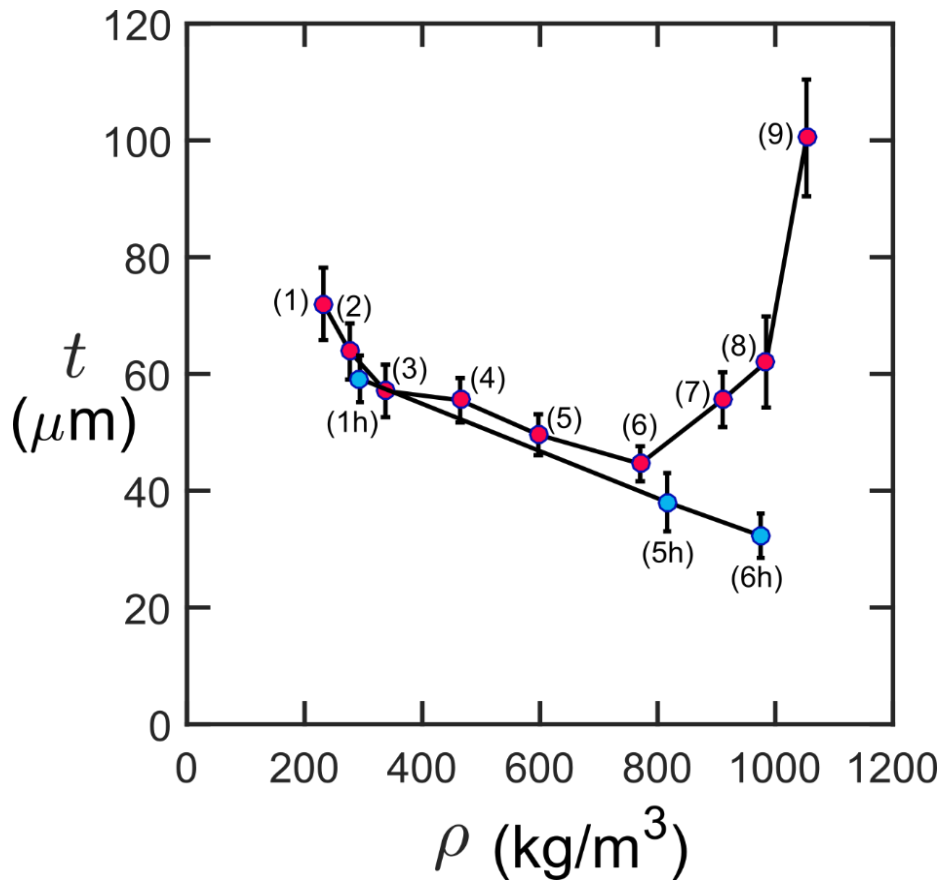

Figure S1: Measured thickness of CNT mat and composites

## 3. Estimation of the axial modulus and yield strength of a CNT bundle

The walls of multi-walled CNTs in the direct-spun mat studied herein are comprised of a number of concentric seamless cylindrical graphene layers, with typical spacing between the layers of $d$ = 0.335 nm in the radial direction [7]. Upon calculating the modulus of a graphene layer based upon an assumed layer thickness of $d$, the longitudinal modulus of the CNT walls according to literature is $E_w = $ 1 TPa [8,9].

The CNT wall density derives from the areal density of carbon atoms in each wall layer $P_S = $ 3.82 x $10^{19}$ m$^{-2}$ [10], the mass of each carbon atom $m_c = 12u$, where the atomic mass unit $u = 1.66 \times 10^{-27}$ kg, and the spacing between each wall layer $d$, as follows:

$$\rho_w = \frac{12 \cdot u \times P_S}{d},$$

where $u = 1.66 \times 10^{-27}$ kg is the atomic mass unit. This yields a wall density $\rho_w = 2.3 \times 10^3$ kg/m$^3$.

In direct-spun CNT materials, measurements of the individual CNTs have shown that they are very long, up to 1 mm in length [11]. Therefore, the CNT lengths are far greater than the length of an individual CNT bundle strut between junctions, which are typically much less than 1 μm. It is therefore reasonable to assume that the CNTs are continuous along the lengths of CNT bundle struts for estimation of the longitudinal bundle modulus.

**4. Geometry of CNT-epoxy unit cells used in simulation**

Images of the unit cells of CNT bundles, epoxy and air used in simulation are given in Figure S2. The ratio of the CNT bundle thickness to the length of unit cell struts $t_B/l$ and the ratio of the epoxy layer thickness to unit cell strut length $t_e/l$ are recorded in Table S2 for each of the simulated unit cells. The ratios of the fillet radii to bundle thickness, $R_1/t_B$ and $R_2/t_B$, are 2.0 and 0.54 respectively, with the exception of composite 6(h), where an elliptical pore was used instead, see Figure S2.

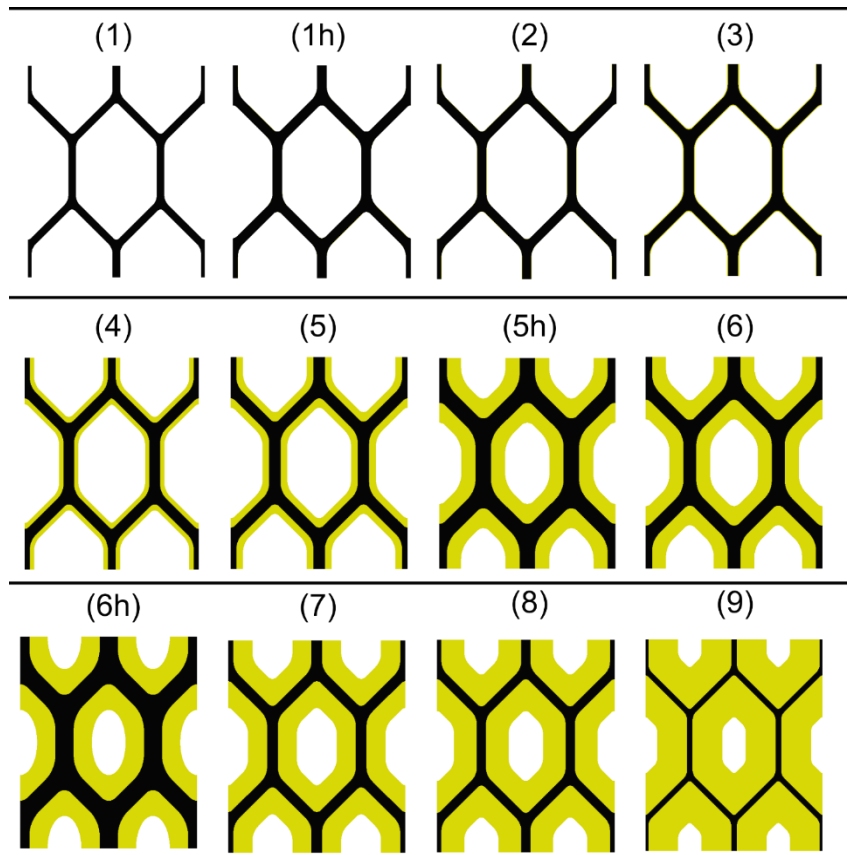

Figure S2: FE models of dry mat and CNT-epoxy composites

Table S2: Normliased values of CNT bundle thickness, epoxy layer thickness and fillet radii of junctions betweeen the struts for FE simulation

| Material | $t_B/l$ | $t_e/l$ |
|---|---|---|
| (1)  | 0.12 | 0 |
| (1h) | 0.16 | 0 |
| (2)  | 0.15 | 0.019 |
| (3)  | 0.17 | 0.013 |
| (4)  | 0.17 | 0.069 |
| (5)  | 0.18 | 0.11 |
| (5h) | 0.25 | 0.23 |
| (6)  | 0.21 | 0.23 |
| (6h) | 0.29 | 0.29 |
| (7)  | 0.17 | 0.31 |
| (8)  | 0.15 | 0.36 |
| (9)  | 0.09 | 0.48 |